# Topological Properties and Functionalities in Oxide Thin Films and Interfaces

Masaki Uchida and Masashi Kawasaki, the University of Tokyo

**Abstract**

As symbolized by the Nobel Prize in Physics 2016, "topology" has been recognized as an essential standpoint to understand and control the physics of condensed matter. This concept may be spreading even into application areas such as novel electronics. In this trend, there has been reported a number of study for the oxide films and heterostructures with topologically non-trivial electronic or magnetic states. In this review, we overview the trends of new topological properties and functionalities in oxide materials with sorting out a number of examples. The technological advances in oxide film growth achieved over the last few decades are now opening the door for harnessing novel topological properties.

## 1. Introduction

Oxide materials have attracted enormous attention because of their unprecedented electronic and magnetic properties and functionalities such as high-temperature superconductivity, colossal magneto- and electro-resistance, and multiferroics. Decades of research have found that these phenomena originate from strong electron correlations in oxides, where charge, spin, and orbital degrees of freedom of the correlated electrons play cooperative roles [1]. Following these understandings, technological advances in epitaxial growth of complex oxides have materialized functional oxide films with equaling or even surpassing properties to those of bulk crystals. The developments have also enabled the growth of metastable oxide phases including various polymorphic structures. In addition, atomically controlled interfaces composed of thoses oxides have been designed and fabricated to realize further emerging phenomena [2]. Their electronic and magnetic behaviors can be controlled by tuning interfacial interactions such as charge transfer, magnetic interactions, and epitaxial strain.

As symbolized by the Nobel Prize in Physics 2016, "topology" has been now recognized as a new important degree of freedom in the field of solid state physics. Topology is a term used to classify electronic or magnetic state, based on its invariant maintained for continuous deformation in a given space. Topological properties are thus essentially robust against perturbations, providing new functionalities possibly suitable for storing and processing information as state variables. Topologically non-trivial states are realized depending on many materials parameters including crystalline symmetry, electron correlations, spin-orbit interactions, and magnetic interactions. In this context, further search for and design of materials are essential to explore possible applications of topological functions [3]. Oxide materials with a wide range of freedom in parameters will be excellent candidates for developing future topological electronics. Indeed, many studies about topological oxide films, interfaces, and superlattices are just beginning of exploration. In this review, we first provide a brief introduction on the general properties and functionalities of topologically non-trivial electronic states and spin structures, and then discuss their new dimensions brought by oxide materials with citing a number of relevant examples.

## 2-1. Properties and functionalities of topological electronic states

A typical example of the topologically non-trivial electronic states appears in the three-dimensional topological insulator [4,5]. Here, conduction and valence band characters are inverted due to strong spin-orbit interactions, giving rise to a helical surface or edge state within the bulk gap, as illustrated in **Fig. 1(a)**. The topological invariant concerned is $Z_2$ invariant (0 or 1), calculated from spatial variances of wave functions in momentum space. The surface state is protected against localization from perturbations (e.g. lattice disorder) which do not break time-reversal symmetry. The surface electronic state has been directly observed such as in bismuth selenides and tellurides by angle-resolved photoemission and scanning tunneling spectroscopies [6-9]. Backward scattering of the surface electron is prohibited due to its spin tightly locked perpendicular to its momentum with an energy scale far above room temperature, which is expected to be applicable in future spintronic devices. In addition, topological insulator with a large bulk gap will enable the use of the dissipationless edge current above room temperature.

More derivatives of topological phases have been reported in recent years. One is a three-dimensional topological Dirac semimetal confirmed such as in transition-metal arsenides [10-13], where conduction and valence bands touch to form gapless bulk Dirac dispersion at the pair of the Dirac points. The two Dirac points are along a rotational axis and protected by corresponding rotational symmetry. The helical surface state called Fermi arc appears to connect the Dirac points. The topological Dirac semimetal is considered as a mother compound for realizing a variety of topological electronic states, such as two- or three- dimensional topological insulator, topological Weyl semimetal, and topological superconductor, by controlling dimensionality, symmetry, and so on. For example, by breaking time-reversal and/or space-inversion symmetry, the Dirac points are further split into two Weyl points and the topological Weyl semimetal phase appears [14-16]. In this case, external fields could control topological phase domains and correspondent dissipationless edge states.

Quantum Hall systems represented by high-mobility gallium arsenide heterostructures are other leading examples of topological electronic states, initiating the research of edge states [17-19]. Recent research is directed toward elucidation of the even-denominator fractional quantum Hall state and possible application of its ground state candidate (Pfaffian state) to quantum computation [20]. Its quasiparticles (Majorana zero modes) obeying non-abelian statistics and their braiding operation are expected to serve as an essential element of topologically protected fault-tolerant quantum computation. In two-dimensional chiral p-wave superconductivity, a topological superconducting state equivalent to the Pfaffian state, the Majorana zero modes are expected to localize in vortex cores. Their non-abelian braiding has also attracted recent interest due to its possible application to the topological quantum computation.

### 2-2. Properties and functionalities of topological spin structures

A striking example of the topologically non-trivial spin structure is magnetic skyrmion [21]. In a typical structure called Bloch-type skyrmion, spins in two dimensions gradually rotate towards the center in perpendicular to the radius directions, as illustrated in **Fig. 1(b)** [22, 23]. The topological invariant concerned is integer skyrmion number (0, ±1, ±2, ...), defined as integral of solid angle subtended by constituent spins in real space. The single skyrmion behaves as a topologically protected spin object, thus attracting rising attention as a robust information carrier in solids. Here, for stabilizing canting of neighboring spins, asymmetric Dzyaloshinskii-Moriya (DM) interaction derived from the spin-orbit coupling is a key parameter. The skyrmion and its lattice structure have been directly observed in chiral metals such as B20-type transition-metal silicides and germanides [22, 24] and a chiral insulator [25] by some special transmission electron microscopy techniques.

Recent studies show promise for the application of the skyrmion to non-volatile magnetic memory, which has the advantages of low driving current and high memory density over the magnetic bubble and racetrack memory devices [26, 27]. The non-coplanar spin alignment in the skyrmion produces emergent magnetic field to conduction electrons, resulting in giant Hall effect that is already applied for detecting the skyrmions [28-33]. Lately, in addition to the Bloch-type skyrmion, more variety of topological spin structures have been discovered such as Néel-type skyrmion [34, 35] and anti-skyrmion [36, 37]. More effort will focus on engineering skyrmionic materials in order to realize stable manipulation of skyrmions above room temperature.

### 3-1. Topological electronic states in oxide films

Compared to selenides and tellurides forming a group of topological insulator materials, reports of three-dimensional topological insulator in oxides have been limited to theoretical suggestions. As listed in **Table 1**, on the other hand, various kinds of derivative topological phases have been predicted and confirmed in oxides. They include novel semimetal phases such as Dirac and Weyl semimetals and further derivative phases. It is immediately obvious that many of them include heavy elements with strong spin-orbit coupling such as iridium, osmium, or bismuth in common. Although electron correlations are rather weak compared to light transition elements, the comparable strength to the spin-orbit interactions has great importance in realizing the novel electronic phases [38].

Iridates have been a rich platform for the research of topological electronic states. The strong spin-orbit interactions lead to low-energy electronic states better represented by effective total angular momentum $J_{\text{eff}}$, and $Ir^{4+}$ ($5d^5$) $t_{2g}$ orbitals are split into $J_{\text{eff}} = 1/2$ and $3/2$ manifolds [39, 40]. Among perovskite iridates, $SrIrO_3$, epitaxially stabilized on perovskite substrates, have attracted growing attention due to its possible exotic topological phases [41-47]. As shown in **Figs. 2(a) and (b)**, correlated semimetallic electronic structures have been clarified by in situ angle-resolved photoemission spectroscopy of the films [41, 42]. The low-energy bands consist of heavy hole-like one at the ($\pi$, 0) point and light electron-like one at the ($\pi/2$, $\pi/2$) point. The remarkable band narrowness is derived not only from strong electron correlations and spin-orbit interactions but also from dimensionality and $IrO_6$ octahedral rotations, which is in stark contrast to less distorted layered perovskite iridates $Sr_2IrO_4$ [39, 40] and $Ba_2IrO_4$ [48, 49]. The steep linear dispersion observed for the electron band (**Fig. 2 (b)**) indicates a formation of Dirac cone in the bulk state, reminiscent of a symmetry-protected nodal line predicted by theoretical calculations [44, 45]. Moreover, it has been theoretically suggested that various topological phase transitions such as to nodal line Weyl semimetal can be induced depending on the direction of the applied magnetic field (**Fig. 2(c)**), by breaking time-reversal symmetry on the degenerate nodal line Fermi surface [45].

Other topological electronic phases, stabilized by magnetic ordering on a pyrochlore lattice, have been intensively studied from both the experimental and theoretical sides [38, 50-79]. Many of pyrochlore iridates $R_2Ir_2O_7$ ($R$=Pr-Lu) show "all-in-all-out" antiferromagnetic spin ordering, where all the four spins pointing inward or outward at the tetrahedron vertices are alternately stacked along the [111] direction. There are two distinct types of magnetic domains (all-in-all-out or all-out-all-in) in this ordering (**Fig. 5(b)**), which are interconnected by time-reversal operation. This antiferromagnetic ordering with the broken time-reversal symmetry is one of the conditions to realize the topological Weyl semimetal (TWS) phase, resulting in odd-parity magnetoresistance depending on the magnetic field direction [61]. While $Pr_2Ir_2O_7$ is metallic down to the lowest temperature, other pyrochlore iridates exhibit a metal-insulator transition accompanied with this all-in-all-out spin ordering [73-79]. Therefore, compounds located on the verge of the metal-insulator transition, such as $Nd_2Ir_2O_7$, are expected to be in or near the TWS phase [72-74], based on theoretical phase diagrams taking into account the spin-orbit interactions and electron correlations [38, 51]. Detailed semimetallic electronic structures reflecting the all-in-all-out spin ordering have been spectroscopically investigated [59, 60, 72, 73], and unique magnetotransport due to its domain-wall or surface state, as explained later, have been also observed [55, 57]. The same all-in-all-out spin ordering has been confirmed for other 5d pyrochlore $Cd_2Os_2O_7$ [64-66, 68], and similar magnetotransport ascribed to the domain-wall conduction in bulk samples has been also reported [67].

A two-dimensional honeycomb lattice has also gained considerable attention in terms of topological phases. $Na_2IrO_3$, composed of the honeycomb lattice of the iridium ions, is one of the leading candidates of a two-dimensional topological insulator or quantum spin Hall insulator state [80, 81]. In addition, as shown in **Figs. 3(a) and (b)**, bilayers of perovskite transition metal oxides have been investigated based on tight-binding modeling and first-principles calculations, because its lattice structure along the [111]-axis can be considered as the honeycomb lattice of the transition metal [82-84]. Topologically non-trivial band gaps in a range of 50-300 meV, which is sufficiently large to realize the quantum spin Hall effect at high temperatures, have been predicted in the bilayer films of perovskite oxides such as $LaAuO_3$ (**Figs. 3(c) and (d)**). Recently, attempts to realize such [111]-oriented heterostructures have been also reported [85, 86]. In these heterostructures, the topological electronic phases are expected to be modulated by applying epitaxial strains or external gate voltages [83].

Anti-perovskite oxides $A_3BO$ ($A$=Ca, Sr, Ba, $B$=Sn, Pb) have attracted growing attention due to their Dirac semimetal state [87-89]. Some bulk experiments with focus on their novel electronic structures and epitaxial growths of their films have begun to be reported [90-93]. Even more simple oxides, rutile oxides such as $IrO_2$, $RuO_2$, and $OsO_2$, have been proposed to host a topological nodal semimetal phase [94]. As more exotic topological phases, an axion insulator state has been theoretically suggested to appear in spinel oxides $CaOs_2O_4$ and $SrOs_2O_4$ [95].

In addition, a topological superconducting state, classified into the two-dimensional chiral *p*-wave symmetry, has long been expected in a layered perovskite $Sr_2RuO_4$ [96-99]. Studies of the unconventional superconductivity will be further advanced using superconducting $Sr_2RuO_4$ thin films, in particular, grown in a reproducible manner by molecular beam epitaxy (MBE) [100] rather than pulsed laser deposition (PLD) [101] (**Fig. 4(a)**). The superconducting $Sr_2RuO_4$ films and junctions will also open a new avenue for detecting and controlling the Majorana zero modes predicted in the vortex cores.

**3-2. Topological electronic states in oxide interfaces**

While formation of the metallic surface state is naturally expected at the heterointerface between topologically trivial and non-trivial electronic states, other type of the surface state is possible in the case of the pyrochlore iridates. Even if the all-in-all-out antiferromagnetic domain itself is not in the TWS phase but in the insulator phase, the surface state should appear at domain walls between all-in-all-out / all-out-all-in domains due to the condition of continuity in the respective bands across the Fermi level (**Fig. 5(b)**) [52, 53]. These states are classified into different topologies from the standpoint of $C_3$ rotational symmetry. In $Nd_2Ir_2O_7$ bulk polycrystals, actually, a huge drop in the resistance has been observed at the field of domain reversal when sweeping the magnetic field, and this can be interpreted as conduction at the magnetic domain walls randomly formed in the reversal process [55, 57]. This domain wall conduction has been clearly demonstrated using a pyrochlore iridate heterointerface (**Fig. 5(a)**) [102]. Here, one layer is $Eu_2Ir_2O_7$, where the domain pattern is fixed by the cooling filed [61]. In the other layer $Tb_2Ir_2O_7$, by contrast, the domain can be reversed by the sweeping field [102]. As a result, as shown in **Fig. 5(b)**, the single domain wall can be created and annihilated at their interface by sweeping the field, and conduction enhancement due to the domain wall conduction is observed with a ferroic hysteresis loop. The pyrochlore iridate heterointerface provides a new stage for investigating the topological surface state, as illustrated in **Fig. 5(c).**

The study of quantum Hall states, other topological electronic phases formed at high-mobility heterointerface, is also being advanced in oxide materials. In ZnO/(Mg,Zn)O heterointerface grown by MBE [103], the highest electron mobility has now exceeded 1,000,000 $cm^2$/Vs [104] and the even-denominator fractional quantum Hall state has been also observed as in the high-mobility GaAs heterointerface [105] (**Fig. 4(b)**). Its ground state will be closely examined from the aspect of the non-abelian Pfaffian state. In ZnO, high controllability of the energy ratio between the Zeeman and Landau splitting is favorable for controlling this fascinating quantum phase. Consideration of the Pfaffian state, such as by measuring quarter electron charge and full spin polarization as in the GaAs heterointerface [106-108], will open new opportunities of application to the topological quantum computation. Other high-mobility two-dimensional electron systems such as $LaAlO_3/SrTiO_3$ heterointerface [109-111] and delta-doped $SrTiO_3$ heterostructures [112-114] can be also the candidates for advancing this direction. The MBE technique has been found to be highly useful for growing high-mobility $SrTiO_3$ films [114-116], and integer quantum Hall effect has been recently observed in the MBE-grown delta-doped heterostructure [112].

**4-1. Topological spin structures in oxide films**

In magnetic skyrmion, as already introduced, its noncoplanar spin alignment is one of the essential characteristics. Unconventional Hall effect observed in such spin configuration has been studied first in a frustrated ferromagnet of pyrochlore $Nd_2Mo_2O_7$ [117], and this is attributed to the finite scalar spin chirality ($\chi_{ijk} = S_i \cdot (S_j \times S_k)$). The resultant Berry phase in real space acts as emergent magnetic field to conduction electrons, and so it is termed as topological Hall effect (THE). THE has been now widely used as an evidence of skyrmion phases in the B20-type silicides and germanides [28-33] and found to be an effective tool for probing skyrmionic structures in various conductive magnetic materials. As listed in **Table 2**, topological spin structures showing THE have been reported also in oxide films and they have unique properties as below.

Classical ferromagnetic semiconductor EuO [118, 119] is one of the leading candidates hosting skyrmionic structures. As shown in **Fig. 6(a)**, topological Hall resistivity $\rho_{THE}$ has been observed as a sharp peak in the magnetization process in EuO, in addition to the conventional anomalous Hall resistivity $\rho_{AHE}$ proportional to magnetization [120-122]. The $\rho_{THE}$ peak appears only in films thinner than $t = 200$ nm. It is rapidly suppressed when tilting the applied magnetic field from the surface normal, indicating the formation of two-dimensional skyrmions. The emergent field $B_e$ (= $\rho_{THE}/R_H$) ~ 0.25 T and the skyrmion radius $r$ (= $t/2 \sin \theta_S$) ~ 140 nm can be also estimated from this behavior, where $R_H$ and $\theta_S$ are the ordinary Hall coefficient and the angle of peak disappearance. This skyrmion is considered to be Bloch-type stabilized in the two-dimensional limit of Heisenberg ferromagnets [123, 124]. A magnetic phase diagram plotting $B_e$ (**Fig. 6(b)**) indicates that the two-dimensional skyrmion phase persists down to the lowest temperature, as in the case of thinned samples of the chiral skyrmionic metals [24] and insulator [25].

Cubic perovskites $SrFeO_3$ and $SrFe_{1-x}Co_xO_3$ are other candidates of skyrmionic oxides [84, 85, 72], which are also located close to metal-insulator transition [1, 128, 129]. Some experiments have suggested that three-dimensional skyrmionic structures are stabilized due to the magnetic frustration, not the DM interaction in this case [125, 126]. The THE signals have been confirmed also in their thin films [127]. While topological properties and dynamics of skyrmions have been studied mostly in metals and insulators, discoveries in the magnetic semiconductors or correlated metals are opening the possibilities of investigating and applying them with controlling the carrier density. In addition to further diffraction measurements or microscopic observations for determining detailed spin textures, experiments of controlling skyrmions for example by gating or photodoping are highly desired. Skyrmionic oxides capable of the Fermi level tuning are also suitable for possible observation of quantized THE, which has been theoretically proposed in high-mobility and low-carrier-density skyrmionic systems [130, 131] such as $K_xRhO_2$ [132].

**4-2. Topological spin structures in oxide interfaces**

Modulation of magnetic exchange interactions in skyrmionic materials is other important direction. In particular, the DM interaction is a key parameter for canting neighboring spins and creating small skyrmions for high memory density. This interaction originates from the inversion symmetry breaking and thus can be spontaneously introduced at heterointerface. In this light, heterostructures consisting of oxide materials with a wide range of magnetic parameters are suitable for realizing size-controlled skyrmion states [133]. Actually, atomic-scale skyrmion lattice [134] and nano-scale individual skyrmions [135] have been verified in various metal multilayers consisting of ferromagnetic metals and strong spin-orbit coupling metals. By varying the composition of the ferromagnetic metal layer, the skyrmion size and density can be also tuned through changes in magnetic interactions [136].

Observation of THE has been reported in such an epitaxial heterostructure of perovskite oxides [137], where one is $SrRuO_3$, a typical itinerant ferromagnet, and the other is $SrIrO_3$, a paramagnetic semimetal with strong spin-orbit coupling (**Fig. 7(a)**). As shown in **Fig. 7(b)**, $\rho_{THE}$ is clearly observed in the Hall measurement of this heterointerface. The $\rho_{THE}$ signal rapidly decreases with increasing thickness of the ferromagnetic $SrRuO_3$ layer and totally disappears above 7 unit cells, strongly indicating that the interfacial DM interaction is essential for stabilizing nano-scale Néel-type skyrmions in the $SrRuO_3$ layer (**Fig. 7(c)**). The results have demonstrated that high-quality oxide heterointerfaces provide a platform for designing topological spin structures. This type of experiment has been spreading into more bilayers of other oxide combinations and even superlattices [138].

**5. Conclusion**

We have overviewed topological electronic states and spin structures in oxide materials with introducing many examples in oxide films and interfaces. Electron correlation in oxides is one of the key parameters for realizing new topological electronic states and spin structures, while the spin-orbit interactions and lattice structures are also important in common. In this sense, complex oxides can be a excellent platform in view of the finely tunable physical parameters by chemical substitution. It should be noted that the thin film fabrication technologies are so advanced in oxides that well-designed heterointerfacee are readily available for promoting further studies of topological electronics. We believe there will be more and more discoveries that explore topological oxide electronics in this frontier field.

**Acknowledgments**

This work was supported by JST CREST Grant No. JPMJCR16F1, Japan and by Grant-in-Aid for Young Scientists (A) No. JP15H05425, and Scientific Research on Innovative Areas "Topological Materials Science" No. JP16H00980 from MEXT, Japan.

Table 1. List of topological electronic states or their candidates in oxide films, interfaces, and superlattices.

| material | topological electronic state | Ref. (theory) | Ref. (experiment) |
|---|---|---|---|
| $SrIrO_3$ | correlated Dirac semimetal (DS) | [43, 44, 45] | [41, 42, 46] |
| $BaBiO_3$, $YBiO_3$ | topological insulator (TI) | [139, 140, 141] | - |
| $Eu_2Ir_2O_7$, $Nd_2Ir_2O_7$ | topological Weyl semimetal (TWS) | [38, 50, 51, 52] | [57, 59, 61, 63] |
| $Bi_2Ir_2O_7$ | TWS | - | [142, 143, 144] |
| $Cd_2Os_2O_7$ | TI or TWS | - | [64, 65, 66, 67, 68] |
| $Tb_2Ir_2O_7/Eu_2Ir_2O_7$ | TWS | [52, 53] | [102] |
| $Na_2IrO_3$ | quantum spin Hall insulator (QSHI) | [80] | [81] |
| $(SrIrO_3)_2$ | QSHI | [82] | [85, 86] |
| $(LaAuO_3)_2$, $(LaNiO_3)_2$ | QSHI | [82, 83, 84] | - |
| $(Y_2Ir_2O_7)_2$ | QSHI | [84, 145] | - |
| $Ca_3SnO$, $Ca_3PbO$ | DS | [87] | [90] |
| $Sr_3SnO$, $Sr_3PbO$ | DS | [88] | [91, 92, 93] |
| $IrO_2$, $RuO_2$, $OsO_2$ | topological nodal line semimetal (TNLS) | [94] | [146, 147, 148] |
| $CaOs_2O_4$, $SrOs_2O_4$ | axion insulator (AI) | [95] | - |

Table 2. List of topological spin structures or their candidates in oxide films, interfaces, and superlattices.

| material | topological spin structure | Ref. (theory) | Ref. (experiment) |
|---|---|---|---|
| $Nd_2Mo_2O_7$ | chirality ordering | - | [117] |
| $Pr_2Ir_2O_7$ | chiral spin liquid | - | [149] |
| $EuO$ | Bloch-type skyrmion | - | [122] |
| $SrFeO_3$, $SrFe_{1-x}Co_xO_3$ | Néel-type skyrmion | - | [125, 126, 127] |
| $Eu_{1-x}Sm_xTiO_3$ | Néel-type skyrmion | - | [150] |
| $SrIrO_3/SrRuO_3$ | Néel-type skyrmion | [137] | [137, 151] |
| $[SrIrO_3/SrRuO_3]_n$ | Néel-type skyrmion | - | [138] |
| $K_xRhO_2$ | quantized topological Hall state | [132] | - |

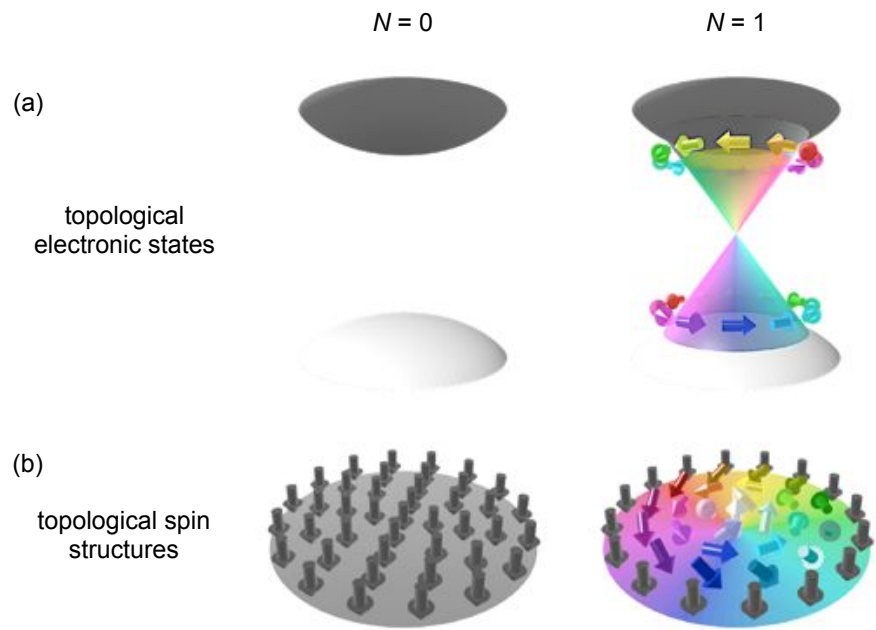

**Figure 1.** Topological (a) electronic states and (b) spin structures characterized by topological invariants *N*, which are defined as integral of variances in the momentum and real spaces, respectively.

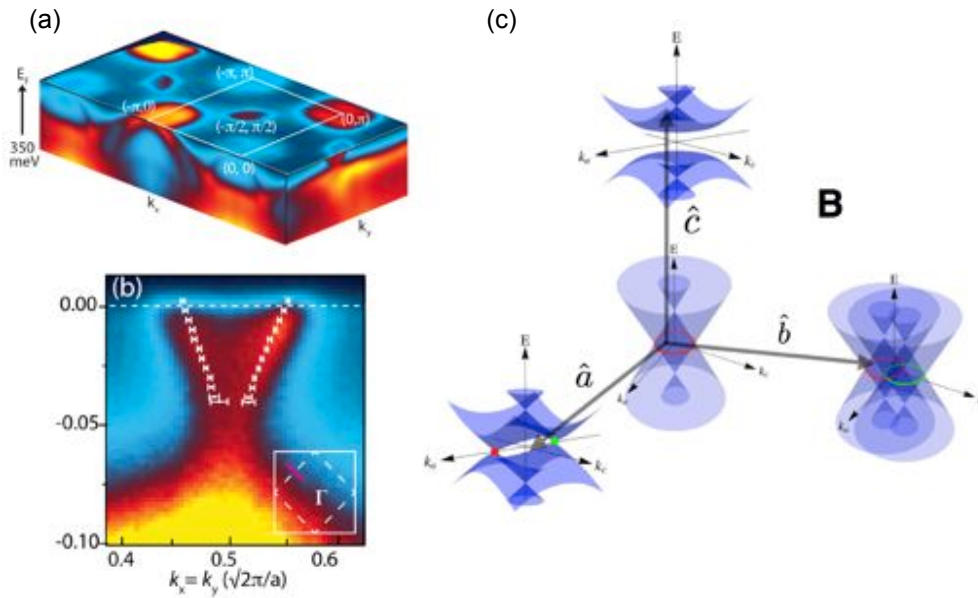

**Figure 2.** (a, b) Energy dispersion of the correlated Dirac semimetal state observed in an epitaxially stabilized $SrIrO_3$ film [41] and (c) various topological phases derived when applying the magnetic field *B* applied to respective directions [45]. Reproduced from Y. F. Nie *et al*., Phys. Rev. Lett. 114, 016401 (2015) and Y. Chen *et al*., Phys. Rev. B 93, 155140 (2016) with the permission of APS journals.

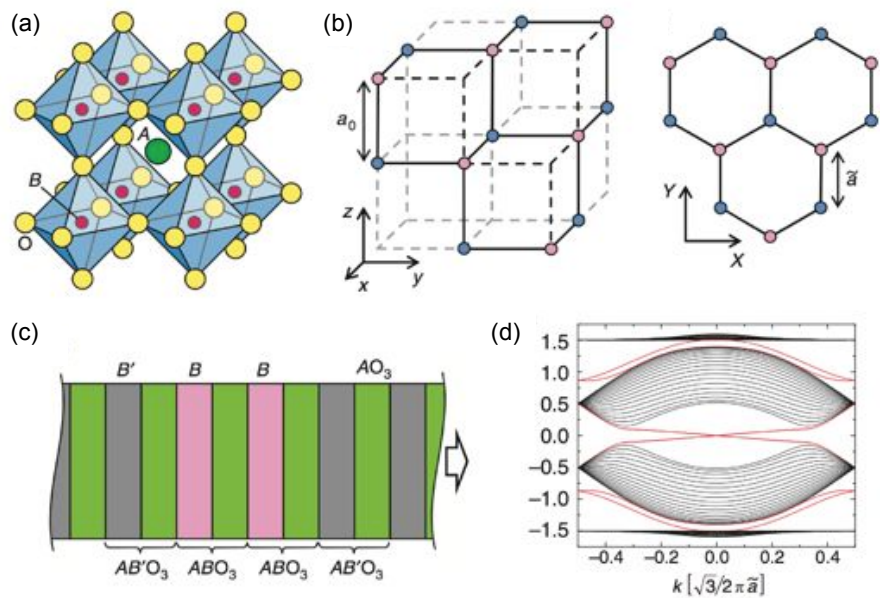

**Figure 3.** (a) Perovskite structure and (b) its bilayer film consisting of the honeycomb lattice in (111) the plane. (c) Confined heterostructure of the perovskite (111) bilayer and (d) topological electronic dispersions calculated for an $e_g$ band model [83]. Reprinted from D. Xiao *et al.*, Nat. Commun. 2, 596 (2011) in accordance with the Creative Commons Attribution (CC BY) license.

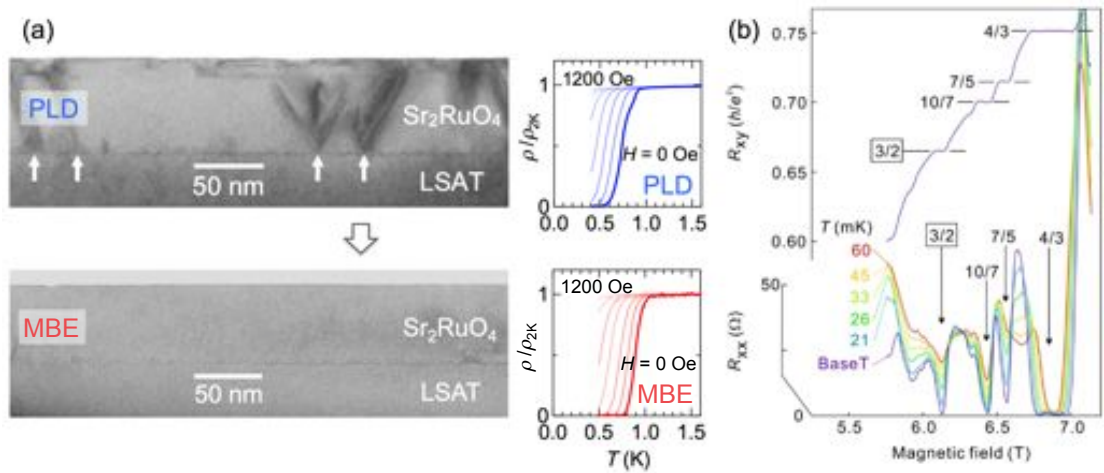

**Figure 4.** (a) Cross-sectional transmission electron microscopy images and superconductivity in $Sr_2RuO_4$ films prepared by PLD and MBE techniques [100, 101]. Reproduced from Y. Krockenberger *et al.*, Appl. Phys. Lett. 97, 082502 (2010) with the permission of AIP Publishing and from M. Uchida *et al.*, APL Mater. 5, 106108 (2017) in accordance with the Creative Commons Attribution (CC BY) license. (b) Even-denominator fractional quantum Hall effect appeared at the filling of 3/2 in ZnO/(Mg,Zn)O heterointerface [105].

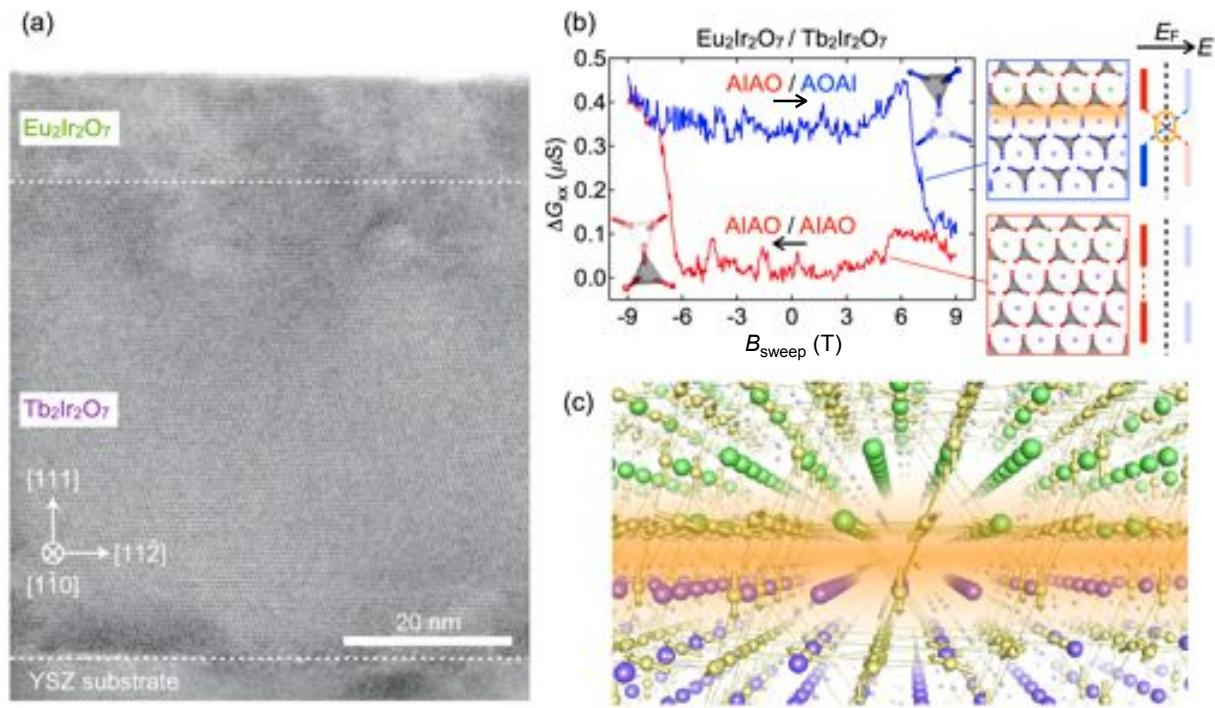

**Figure 5.** (a) Atomically resolved transmission electron microscopy image of the $Eu_2Ir_2O_7$/$Tb_2Ir_2O_7$ pyrochlore heterointerface. (b) Metallic conduction observed in the single domain wall plane at the heterointerface between all-in-all-out (AIAO) / all-out-all-in (AOAI) magnetic domains [102] as depicted in (c) its schematic with atomic arrangement. Reproduced from T. C. Fujita *et al.*, Phys. Rev. B 93, 064419 (2016) with the permission of APS journals.

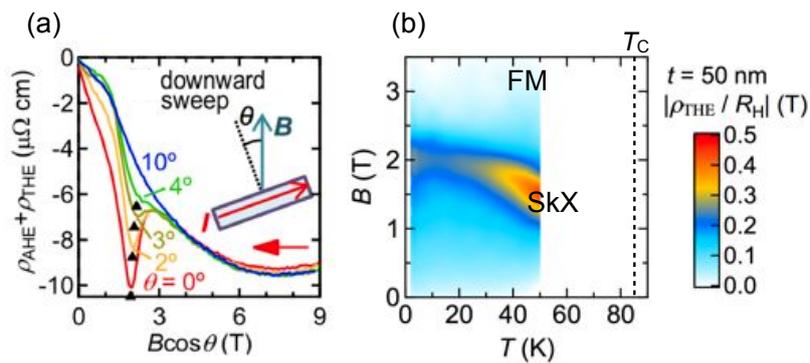

**Figure 6.** (a) Topological Hall resistivity observed for EuO film ($t = 50$ nm) and its field-angle dependence. (b) Magnetic phase diagram constructed by mapping the normalized intensity of the topological Hall resistivity [122]. Reproduced from Y. Ohuchi *et al.*, Phys. Rev. B 91, 245115 (2015) with the permission of APS journals.

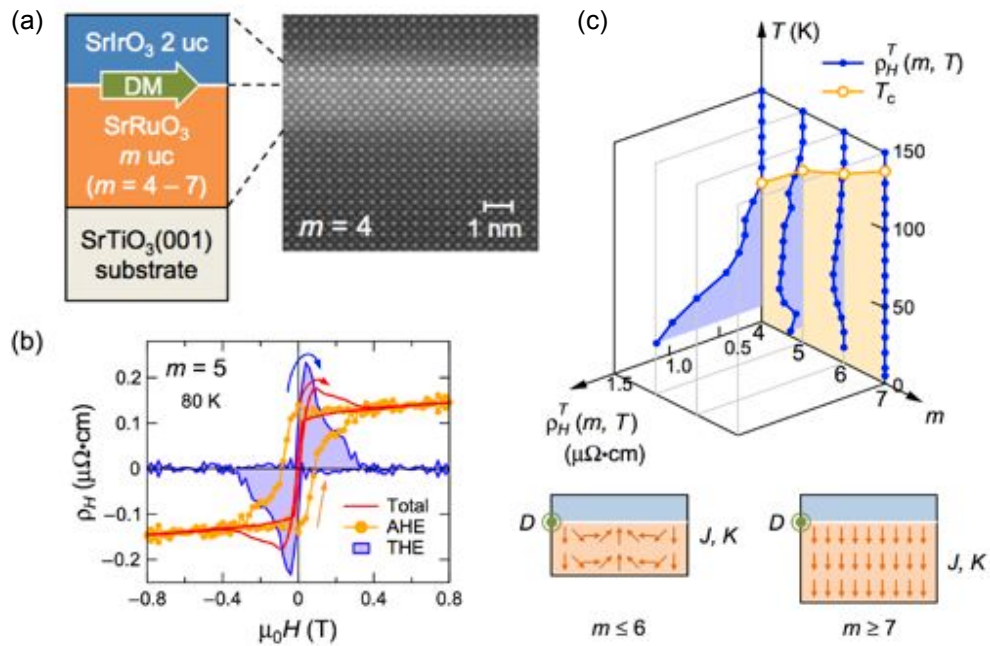

**Figure 7.** (a) The cross-sectional schematic and atomically resolved transmission electron microscopy image for the SrIrO$_3$/SrRuO$_3$ perovskite heterointerface. (b) Topological Hall resistivity (blue) deduced by the subtraction of anomalous Hall contribution (orange) from measured Hall signal (red). (c) Stability of skyrmions as a function of temperature and SrRuO$_3$ thickness. As SrRuO$_3$ gets thicker, ferromagnetic interaction ($J$) surpass DM interaction ($D$) [137]. Reprinted from J. Matsuno *et al.*, Sci. Adv. 2, e1600304 (2016) in accordance with the Creative Commons Attribution (CC BY) license.